\PassOptionsToPackage{dvipsnames}{xcolor}
\PassOptionsToPackage{htt}{hyphenat}

\documentclass[acmsmall]{acmart}

\AtBeginDocument{%
}

\setcopyright{acmlicensed}
\copyrightyear{2026}
\acmYear{2026}
\acmDOI{XXXXXXX.XXXXXXX}

\acmJournal{JACM}
\acmVolume{37}
\acmNumber{4}
\acmArticle{111}
\acmMonth{7}

\usepackage{booktabs}
\usepackage{siunitx}
\usepackage[dvipsnames]{xcolor}
\usepackage{tikz}
\usepackage{amsmath}
\usepackage{algpseudocode}
\usepackage[english]{babel}
\usepackage[utf8]{inputenc}
\usepackage{hyphenat}
\usepackage{algorithm}

\usepackage{svg}
\usepackage{enumitem}
\usepackage{minted}
\usepackage{soul}
\usepackage{balance}
\usepackage{multicol}
\usepackage{bbding}
\usepackage{svg}
\usepackage{ifthen}
\usepackage{csquotes}
\usepackage{float}
\usepackage[T1]{fontenc}
\usepackage{mathrsfs}
\usepackage{pifont}
\usepackage{booktabs}
\usepackage{array}
\usepackage{booktabs}
\usepackage{multirow}
\usepackage{tabularx}
\usepackage{xcolor}
\usepackage{multirow} 
\usepackage{listings}    
\usepackage{tasks}

\usepackage{graphicx}    
\usepackage{lipsum}      
\usepackage{tikz} 
\usepackage[most]{tcolorbox}
\usepackage{subcaption}

\usepackage{pgfplots}
\usepackage{xcolor}
\pgfplotsset{compat=1.18}
\usepgfplotslibrary{groupplots} 

\usepackage{pgfplotstable}
\usepackage{longtable}
\usepackage{subcaption}
\usepackage{booktabs}   
\usepackage{tabularx}   
\usepackage{ragged2e}   
\usepackage[T1]{fontenc}
\usepackage[htt]{hyphenat}  

\definecolor{javablue}{RGB}{0, 51, 153}     
\definecolor{javagreen}{RGB}{0, 85, 0}       
\definecolor{javapurple}{RGB}{102, 51, 153}  
\definecolor{javagray}{RGB}{90, 90, 90}      
\definecolor{javapunct}{RGB}{50, 50, 50}     
\definecolor{javabg}{RGB}{252, 252, 252}     
\definecolor{javaborder}{RGB}{160, 160, 160} 

\definecolor{backcolour}{RGB}{249,249,251}
\definecolor{codegray}{RGB}{110,115,125}
\definecolor{codeborder}{RGB}{220,223,230}
\definecolor{punct}{RGB}{140,140,150}
\definecolor{delim}{RGB}{120,120,135}
\definecolor{accent}{RGB}{0,95,160}

\definecolor{codegreen}{rgb}{0,0.6,0}
\definecolor{codegray}{rgb}{0.5,0.5,0.5}
\definecolor{codepurple}{rgb}{0.58,0,0.82}
\definecolor{backcolour}{rgb}{0.95,0.95,0.95}
\definecolor{coolblue}{RGB}{145,191,219}

\newtcolorbox{rqbox}[1][]{
  colback=gray!5!white,
  colframe=gray!50!black,
  fonttitle=\bfseries,
  title=#1,
  title after break={},
  boxrule=0.5pt,
  arc=2pt,
  left=6pt,
  right=6pt,
  top=4pt,
  bottom=4pt
}

\lstdefinestyle{JavaStyle}{
  language=Java,
  backgroundcolor=\color{javabg},
  basicstyle=\ttfamily\footnotesize,   
  numbers=left,
  numberstyle=\scriptsize\color{javagray},
  numbersep=10pt,
  firstnumber=1,
  stepnumber=1,
  frame=single,
  rulecolor=\color{javaborder},
  framerule=0.4pt,
  framesep=7pt,
  xleftmargin=1.25em,
  xrightmargin=0.5em,
  framexleftmargin=1.25em,
  keywordstyle=\color{javablue}\bfseries,
  commentstyle=\color{javagreen}\itshape,
  stringstyle=\color{javapurple},
  breaklines=true,
  breakatwhitespace=true,
  breakindent=1.25em,
  prebreak=\raisebox{0ex}[0ex][0ex]{\tiny\(\hookleftarrow\)},
  postbreak=\mbox{\textcolor{javagray}{\tiny\(\rightarrow\)}\space},
  showspaces=false,
  showtabs=false,
  showstringspaces=false,
  keepspaces=true,
  columns=fullflexible,
  tabsize=2,
  upquote=true,
  captionpos=b,
  literate=
    {:}{{{\color{javapunct}{:}}}}1
    {,}{{{\color{javapunct}{,}}}}1
    {;}{{{\color{javapunct}{;}}}}1
    {\{}{{{\color{javapunct}{\{}}}}1
    {\}}{{{\color{javapunct}{\}}}}}1
    {(}{{{\color{javapunct}{(}}}}1
    {)}{{{\color{javapunct}{)}}}}1
}

\lstnewenvironment{javacode}[1][]%
  {\lstset{style=JavaStyle,#1}}{}

\lstdefinelanguage{json}{
  showstringspaces=false,
  breaklines=true,
  stringstyle=\color{codepurple},
  morestring=[s]{"}{"},
  literate=
   *{0}{{{\color{blue}0}}}{1}
    {1}{{{\color{blue}1}}}{1}
    {2}{{{\color{blue}2}}}{1}
    {3}{{{\color{blue}3}}}{1}
    {4}{{{\color{blue}4}}}{1}
    {5}{{{\color{blue}5}}}{1}
    {6}{{{\color{blue}6}}}{1}
    {7}{{{\color{blue}7}}}{1}
    {8}{{{\color{blue}8}}}{1}
    {9}{{{\color{blue}9}}}{1}
    {:}{{{\color{javapunct}{:}}}}{1}
    {,}{{{\color{javapunct}{,}}}}{1}
    {\{}{{{\color{javapunct}{\{}}}}{1}
    {\}}{{{\color{javapunct}{\}}}}}{1}
    {[}{{{\color{javapunct}{[}}}}{1}
    {]}{{{\color{javapunct}{]}}}}{1}
    {true}{{{\color{javagreen}true}}}{4}
    {false}{{{\color{javagreen}false}}}{5}
    {null}{{{\color{javagreen}null}}}{4},
}

\lstdefinestyle{JSONStyle}{
  language=json,
  backgroundcolor=\color{javabg},
  basicstyle=\ttfamily\footnotesize,   
  numbers=left,
  numberstyle=\scriptsize\color{javagray},
  numbersep=10pt,
  firstnumber=1,
  stepnumber=1,
  frame=single,
  rulecolor=\color{javaborder},
  framerule=0.4pt,
  framesep=7pt,
  xleftmargin=1.25em,
  xrightmargin=0.5em,
  framexleftmargin=1.25em,
  breaklines=true,
  breakatwhitespace=true,
  breakindent=1.25em,
  prebreak=\raisebox{0ex}[0ex][0ex]{\tiny\(\hookleftarrow\)},
  postbreak=\mbox{\textcolor{javagray}{\tiny\(\rightarrow\)}\space},
  moredelim=**[is][\color{javagreen}]{@}{:},
  showstringspaces=false,
  showtabs=false,
  keepspaces=true,
  columns=fullflexible,
  captionpos=b,
  tabsize=2,
  upquote=true
}

\lstnewenvironment{jsoncode}[1][]%
  {\lstset{style=JSONStyle,#1}}{}

\usepackage{upquote} 

\lstdefinestyle{PromptStyle}{
  backgroundcolor=\color{backcolour},
  basicstyle=\ttfamily\footnotesize,   
  numbers=left,
  numberstyle=\scriptsize\color{codegray},
  numbersep=10pt,
  firstnumber=1,
  stepnumber=1,
  frame=single,
  rulecolor=\color{codeborder},
  framerule=0.4pt,
  framesep=7pt,
  xleftmargin=1.25em,
  xrightmargin=0.5em,
  framexleftmargin=1.25em,
  breaklines=true,
  breakatwhitespace=true,              
  breakindent=1.25em,
  prebreak=\raisebox{0ex}[0ex][0ex]{\tiny\(\hookleftarrow\)},
  postbreak=\mbox{\textcolor{codegray}{\tiny\(\rightarrow\)}\space},
  showspaces=false,
  showtabs=false,
  showstringspaces=false,
  keepspaces=true,
  columns=fullflexible,
  tabsize=2,
  upquote=true,
  captionpos=b,
  escapeinside={(*@}{@*)},
  literate=
    {“}{{"}}1 {”}{{"}}1 {’}{{'}}1
    {–}{{-}}1 {—}{{-}}1
    {:}{{{\color{punct}{:}}}}1
    {,}{{{\color{punct}{,}}}}1
    {\{}{{{\color{delim}{\{}}}}1
    {\}}{{{\color{delim}{\}}}}}1
    {[}{{{\color{delim}{[}}}}1
    {]}{{{\color{delim}{]}}}}1
}

\lstnewenvironment{promptcode}[1][]%
  {\lstset{style=PromptStyle,#1}}{}

\newcommand{\zf}{{\sc ZeroFalse}}
\newcommand{\ov}{{OpenVuln}}

\definecolor{commentcolor}{RGB}{63, 127, 95}
\definecolor{functioncolor}{RGB}{0, 0, 128}
\definecolor{keywordcolor}{RGB}{128, 0, 0}
\definecolor{builtin}{RGB}{51, 51, 255}
\newminted{yaml}{frame=lines,framerule=2pt}
\newboolean{includeappendix}
\setboolean{includeappendix}{true} 
\begin{document}

\title{ZeroFalse: Improving Precision in Static Analysis with LLMs}


\author{Mohsen Iranmanesh}
\affiliation{%
  \institution{Simon Fraser University}
  \city{Burnaby}
  \country{Canada}}
\email{mia32@sfu.ca}

\author{Sina Moradi Sabet}
\affiliation{%
  \institution{Amirkabir University of Technology}
  \city{Tehran}
  \country{Iran}}
\email{sina.moradi@aut.ac.ir}

\author{Sina Marefat}
\affiliation{%
  \institution{K. N. Toosi University of Technology}
  \city{Tehran}
  \country{Iran}}
\email{sina.marefat@email.kntu.ac.ir}

\author{Ali Javidi Ghasr}
\affiliation{%
  \institution{Ferdowsi University of Mashhad}
  \city{Mashhad}
  \country{Iran}}
\email{alijavidighasr@mail.um.ac.ir}

\author{Allison Wilson}
\affiliation{%
  \institution{Cyber Risk Solutions}
  \city{}
  \country{Canada}}
\email{awilson@cyberrisk-solutions.com}

\author{Iman Sharafaldin}
\affiliation{%
  \institution{Forward Security}
  \city{Vancouver}
  \country{Canada}}
\email{i.sharafaldin@fwdsec.com}

\author{Mohammad A. Tayebi}
\affiliation{%
  \institution{Simon Fraser University}
  \city{Burnaby}
  \country{Canada}}
\email{tayebi@sfu.ca}

\renewcommand{\shortauthors}{Iranmanesh et al.}

\begin{abstract}
Static Application Security Testing (SAST) tools are integral to modern software development, yet their adoption is undermined by excessive false positives that weaken developer trust and demand costly manual triage. We present \zf, a framework that integrates static analysis with large language models (LLMs) to reduce false positives while preserving coverage. \zf\ treats static analyzer outputs as structured contracts, enriching them with flow-sensitive traces, contextual evidence, and CWE-specific knowledge before adjudication by an LLM. This design preserves the systematic reach of static analysis while leveraging the reasoning capabilities of LLMs. We evaluate \zf\ across both benchmarks and real-world projects using ten state-of-the-art LLMs. Our best-performing models achieve F1-scores of 0.912 on the OWASP Java Benchmark and 0.955 on the \ov~ dataset, maintaining recall and precision above 90\%. Results further show that CWE-specialized prompting consistently outperforms generic prompts, and reasoning-oriented LLMs provide the most reliable precision–recall balance. These findings position \zf\ as a practical and scalable approach for enhancing the reliability of SAST and supporting its integration into real-world CI/CD pipelines.

\end{abstract}






\maketitle

\section{Introduction}
 Static Application Security Testing (SAST), is a foundational pillar of the modern Software Development Lifecycle (SDLC) \cite{1628970}. By analyzing source code prior to execution, SAST tools aim to detect security vulnerabilities early, reducing remediation costs and preventing critical defects from reaching production \cite{mcgraw2004software}. Integrating automated security analysis directly into CI/CD pipelines is essential to maintain development pace while building secure software \cite{10621730,10034883}.

However, the practical efficacy of SAST is severely undermined by a high rate of false positives \cite{10.1145/3338112}. This sheer number of incorrect or irrelevant warnings overwhelms developers, causing ``alert fatigue'' and leading to a lack of trust in the tools \cite{johnson2013why, sadowski2018tricorder,8987507,6606613}. Consequently, valid alerts are often ignored, allowing genuine vulnerabilities to persist. While traditional mitigation techniques like rule-set tuning, combining dynamic analysis with static analysis \cite{busse2022combining}, fuzzing \cite{wustholz2020targeted, 10179296, wen2020memlock}, and conventional machine learning classifiers exist \cite{4815348,8730202,10.1145/2597073.2597100}, they often lack the deep semantic understanding required to differentiate between a true vulnerability and a benign coding pattern, proving insufficient to solve the problem at scale \cite{heckman2011systematic}.

The advent of LLMs presents a transformative opportunity to overcome this long-standing challenge. With a powerful capacity to comprehend and reason about both natural language and source code \cite{hou2024large}, LLMs can analyze the rich context surrounding a potential vulnerability, from developer comments to complex data flows, in a manner previously unattainable for automated systems \cite{chen2021evaluating,nam2024using}. While there is a growing body of work on applying LLMs to vulnerability detection \cite{keltek2025lsast, liu2024exploration, yin2024multitask, li2023assisting, yang2025knighter}, research on hybrid solutions that use LLMs to improve static analysis results remains limited. LLM4SA~\cite{wen2024automatically} was the first LLM-based approach to mitigating false positives in static analysis. Building on this, LLM4FPM~\cite{chen2024utilizing} applies code slicing to provide LLMs with concise yet complete code contexts for each alert. Although both frameworks showed promise, their evaluations were limited to a small set of CWEs and did not explore differences between vulnerability types. Furthermore, neither study systematically compared the performance of different LLMs, whether open-source or commercial. These limitations motivated our work, which expands CWE coverage and benchmarks multiple state-of-the-art LLMs under a unified evaluation setting.

In this paper, we introduce \zf\  a novel framework that leverages LLMs to enhance the precision of static analysis by intelligently filtering false positives. Our approach is grounded in a key hypothesis: a ``one-size-fits-all'' approach is suboptimal for the diverse landscape of vulnerabilities. Instead, \zf\ is designed to take advantage of CWE-specific optimization, tailoring the analysis to the unique characteristics of different vulnerability classes (e.g., SQL Injection vs. Cross-Site Scripting) to achieve superior accuracy. Our framework systematically provides the LLM with crucial context, including not only code snippets from the functions containing the alert but also the precise dataflow path identified by the SAST tool, enriched with CWE-aware metadata (such as the type of vulnerability, CWE-specific micro-rubrics, boundary checklist, and context-specific patterns). This CWE-driven context enables the LLM to reason about vulnerabilities in a more specialized manner, thereby effectively reducing false positives.

To assess our approach, we conduct a large-scale empirical study of \zf\ using ten state-of-the-art LLMs spanning major families, including Google Gemini, OpenAI GPT, xAI Grok, Mistral, DeepSeek, and Qwen. This diversity ensures that our findings are not biased toward a single architecture or training paradigm (e.g., dense vs.\ mixture-of-experts). We evaluate \zf\ on alerts generated by CodeQL over two datasets: the OWASP Java Benchmark~\cite{owaspbenchmark} and \ov, our curated dataset of real-world alerts from seven widely used open-source projects.  

Across both datasets, \zf\ consistently improves false-positive mitigation while maintaining high recall. On OWASP, models such as \texttt{grok-4} (F1=0.912) and \texttt{gemini-2.5-pro} (F1=0.910) achieved strong precision–recall trade-offs, while \texttt{gpt-oss-20b} delivered the highest recall (0.900). On the more challenging \ov~ dataset, \texttt{gpt-5} emerged as the clear leader (F1=0.955, Recall=0.9142), followed by \texttt{grok-4} (F1=0.923) and \texttt{gpt-oss-20b} (F1=0.904). In contrast, some large-scale models, such as \texttt{gemini-2.5-pro}, collapsed under real-world noise (F1=0.372). These results confirm that reasoning-oriented LLMs, supported by CWE-specific prompting, provide the most reliable false-positive reduction, while raw scale or extended context windows alone do not guarantee generalization. 

{In summary, this paper makes the following contributions: $1)$ \zf\ framework: a principled integration of static analysis with LLM-based adjudication, leveraging structured evidence, CWE-aware prompting, and deterministic outputs; $2)$ Comprehensive evaluation: a large-scale study of LLMs for false-positive reduction in static analysis, spanning multiple datasets, and covering a diverse set of nine widely used LLMs; and $3)$ Empirical insights: evidence that CWE-specialization and reasoning-oriented architectures deliver substantial improvements in precision without compromising recall.}

Our findings show that combining the strengths of static analysis and LLMs transforms SAST from a noisy, low-trust signal into a precise and actionable security tool. We believe \zf\ marks a step toward restoring developer confidence in automated security analysis and making it a practical pillar of secure software development at scale.

\section{Background}

\subsection{Static Analysis and Limitations}

SAST tools prioritize soundness over precision by relying on conservative approximations of program behavior \cite{gens2018k, machiry2017dr}. To minimize false negatives, they often assume that all code paths are feasible, flagging any potential dataflow from untrusted sources to sensitive sinks \cite{blackshear2018racerd}. This approach, although effective in ensuring that every potentially vulnerable path is reported so that no true vulnerability is overlooked, produces frequent false positives due to several recurring factors: over-approximation of logically infeasible paths, limited semantic understanding of developer intent and custom sanitization, context-insensitive analysis across loops, branches, and interprocedural calls, and incomplete or inaccurate models of library or helper functions \cite{10.1145/3338112}. To illustrate these challenges in practice, we present an example from OWASP.

Below is a code snippet from OWASP that illustrates this issue. CodeQL flags a potential SQL injection vulnerability because the untrusted input \lstinline[style=JavaStyle]|param| flows into an SQL query string \lstinline[style=JavaStyle]|sql|. However, the program logic ensures that the value of the variable \lstinline[style=JavaStyle]|bar| is deterministically safe. The \lstinline[style=JavaStyle]|param| value is inserted into a list, but after removing the first element, the final assignment retrieves a controlled list element, resulting in \lstinline[style=JavaStyle]|"moresafe"| being selected. Despite this logic, the static analyzer only observes that \lstinline[style=JavaStyle]|param| (a tainted source) reaches the SQL sink through \lstinline[style=JavaStyle]|bar|. Without deep semantic reasoning about list operations, the analyzer produces a false positive. This example highlights the trade-off inherent in static analysis: maximizing coverage often requires conservative assumptions that sacrifice precision.

\begin{lstlisting}[style=JavaStyle
    % caption={A false positive SQL injection (CWE-089) in OWASP BenchmarkTest00205.},
    % label={fig:java_code}
    ]
String param = "";
if (request.getHeader("BenchmarkTest00205") != null) {
    param = request.getHeader("BenchmarkTest00205");
}
param = java.net.URLDecoder.decode(param, "UTF-8");

String bar = "alsosafe";
if (param != null) {
    java.util.List<String> valuesList = new java.util.ArrayList<String>();
    valuesList.add("safe");
    valuesList.add(param);
    valuesList.add("moresafe");

    valuesList.remove(0); // remove the 1st safe value
    bar = valuesList.get(1); // get the last 'safe' value
}

String sql = "INSERT INTO users (username, password) VALUES ('foo','" + bar + "')";

\end{lstlisting}

\subsection{False Positive Reduction}

The problem of reducing false positives in static analysis has evolved significantly over the past decades, progressing from simple heuristic filters to sophisticated machine learning techniques \cite{10.1145/3510003.3510153,8730202,10.1145/1349332.1349339,liu2018mining}. Nevertheless, achieving high precision without compromising recall remains a persistent challenge in software security. Traditional approaches typically involve manual rule refinement to tune sensitivity for individual projects \cite{mangal2015user}, ranking systems that prioritize alerts based on historical patterns \cite{kim2007warnings,kim2007prioritizing}, or supervised classifiers trained on labeled datasets distinguishing true from spurious alerts \cite{yuksel2013automated, yoon2014reducing, zhang2020variable}. Despite these efforts, such methods often fail to generalize across diverse codebases and struggle to capture the subtle semantic distinctions that separate genuine vulnerabilities from benign coding practices. 

Recent advances in LLMs have introduced a transformative paradigm for false positive reduction. Leveraging their deep semantic reasoning capabilities, LLMs can analyze code context, data flows, and security mechanisms in ways unattainable by conventional techniques. By integrating structured outputs from static analyzers with LLM-based reasoning, these hybrid approaches substantially improve precision while preserving high recall. A crucial enabler of this integration is the Static Analysis Results Interchange Format (SARIF), which has become the de facto industry standard for representing static analysis findings in a structured and machine-readable manner \cite{SARIF-v2.1.0-Errata01}. In addition to specifying issue locations and categories, SARIF provides fields for severity levels, CWE mappings, explanatory messages, and dataflow traces from source to sink. While support for these fields varies across tools, CodeQL in particular populates them comprehensively, making it well-suited for our framework. This rich representation not only facilitates interoperability across diverse analysis tools but also provides a consistent and extensible foundation for higher-level reasoning. Within our framework, SARIF functions as the central interface between traditional static analyzers and LLM-based adjudication, preserving the logical chains necessary for vulnerability identification while enabling automation, contextual enrichment, and precise evidence presentation.

\subsection{CWE Categories}

CWE, maintained by MITRE, is a community-driven taxonomy of software weakness types \cite{cwemitre2025}. Unlike Common Vulnerabilities and Exposures (CVE) \cite{CVEmitre2025}, which catalogs specific vulnerability instances, CWE provides a standardized vocabulary for recurring structural flaws in code. This taxonomy is widely adopted in secure software development, academic research, and benchmarking.

In this study, we focus on CWE categories that capture common and high-impact security flaws across web and systems software. Injection-related weaknesses (CWE-078/089/090/643) arise when untrusted input is incorporated into commands or queries; they include command injection, SQL injection, LDAP injection, format string vulnerabilities, and XPath injection. Web-specific weaknesses cover cross-site scripting (CWE-079), which threaten web application integrity and user trust. Path traversal (CWE-022) enables unauthorized file access through crafted input. Data protection weaknesses (CWE-327) involve insecure handling of sensitive data, such as use of obsolete cryptographic algorithms. Finally, trust boundary violations (CWE-501) capture cases where implicit assumptions about privilege separation are broken, requiring contextual reasoning about data and authority flows.

Organizing evaluation results by CWE categories provides fine-grained insight into capabilities of detection models. It highlights which classes of weaknesses benefit from pattern recognition (e.g., injection flaws with strict syntax) and which require deeper contextual reasoning (e.g., sanitization or trust boundaries). At the same time, CWE-based benchmarking has limitations: categories vary in granularity, and benchmark coverage is uneven across weaknesses. Results should therefore be interpreted as relative indicators of model strengths rather than exhaustive coverage of the security landscape.

\subsection{Characterizing LLMs for Code Analysis}

LLMs have demonstrated remarkable capabilities in code understanding, capturing programming idioms, security practices, and contextual signals that extend well beyond the scope of traditional rule-based static analysis\cite{fan2023large, zhou2025large}. Unlike conventional tools that rely on predefined checks, LLMs leverage auxiliary cues such as comments, variable names, and documentation to reason about code semantics. This enables the detection of framework-specific features, sanitization techniques, and subtle dataflow risks that are often missed by static analyzers. Recent advances from OpenAI, Google, and others underscore the ability of modern LLMs to track tainted data, identify recurring vulnerability patterns, and reason across long, interprocedural code segments.  

To explain the performance variations observed in our experiments, we classify the evaluated models along three axes: architectural density (dense vs. sparse), advanced reasoning frameworks, and model size. These dimensions reflect an evolving design space in which LLMs increasingly combine multiple paradigms to balance efficiency, accuracy, and reasoning depth.  

\vspace{1mm}
\noindent \textbf{Dense vs. Sparse Architectures (Mixture-of-Experts).} While most LLMs adopt the transformer architecture, they differ in parameter utilization. Dense models (e.g., DeepSeek: R1 Distill Llama 70B \cite{guo2025deepseek}) activate the full parameter set for every token, maximizing capacity at the cost of efficiency. In contrast, sparse Mixture-of-Experts (MoE) models (e.g., Mixtral \cite{jiang2024mixtral}, Gemini 2.5 \cite{comanici2025gemini}, DeepSeek \cite{guo2025deepseek}, Qwen3-235B \cite{yang2025qwen3}) selectively route tokens to specialized experts, reducing computational overhead while enabling targeted specialization for tasks such as vulnerability detection. 

\vspace{1mm}

\noindent \textbf{Advanced Reasoning Frameworks.}
Recent models embed mechanisms to improve multi-step reasoning. Router-based systems (e.g., GPT-5) dynamically dispatch queries to fast or deep models. Multi-mode systems (Gemini 2.5, DeepSeek R1, Qwen3-235B) allow extended “thinking” modes. Native chain-of-thought models (OpenAI o4-mini) explicitly structure reasoning steps. Reinforcement learning approaches (DeepSeek-R1) directly train reasoning capabilities. These advances improve the ability of models to separate true vulnerabilities from false positives.

\vspace{1mm}

\noindent \textbf{Model Size.}
Model size remains a key factor in determining capability. Larger models, with hundreds of billions of parameters (e.g., Qwen3-235B, GPT-5), capture broader code and security patterns, improving semantic reasoning and vulnerability detection. However, smaller models (e.g., Mixtral 8x7B, GPT OSS20B, DeepSeek R1 Distill-70B) can achieve competitive performance when paired with effective prompting or domain-specific fine-tuning, making them suitable for cost-sensitive or latency-constrained CI/CD environments. The trade-off between size, accuracy, and deployment efficiency remains central to practical adoption.  

\section{Methodology}

\begin{figure*}[t]
    \centering
    \includegraphics[width=\textwidth]{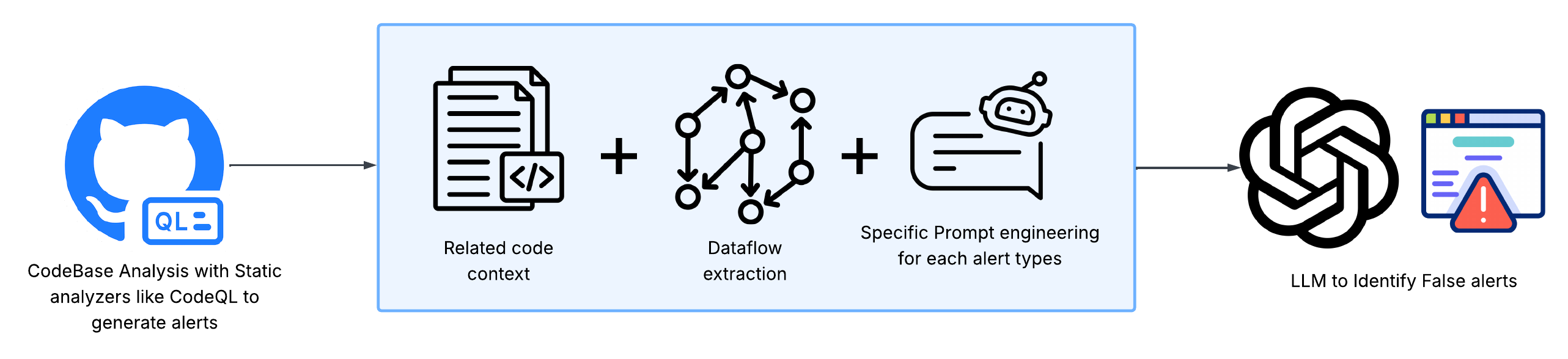}
    \caption{The \zf\ ~pipeline. CodeQL generates alerts, related code context is collected, the dataflow is extracted and annotated, CWE-specific prompts are constructed, and finally, an LLM classifies alerts to identify false positives.}
    \label{fig:zerofalse_pipeline}
    \Description{A flowchart with five stages labeled A through E. Stage A is CodeQL analysis. Stage B is Context Harvesting. Stage C is Dataflow Extraction. Stage D is Prompt Construction. Stage E is LLM Adjudication, which produces the final labeled findings.}
\end{figure*}

Our proposed solution for FP reduction, \zf{}, combines the strengths of static analysis and LLMs to reduce false positives in vulnerability detection. It preserves the broad coverage of static analysis while leveraging LLM-based classification to filter out spurious findings. The central insight is to treat SARIF alerts as structured contracts that can be systematically enriched with contextual evidence, flow-sensitive annotations, and CWE-specific knowledge before being passed to an LLM for adjudication. To ensure reliability, \zf\ employs deterministic prompts, schema-constrained outputs, and a complete audit trail, enabling reproducible and trustworthy false-positive suppression at scale. Figure~\ref{fig:zerofalse_pipeline} depicts the \zf\  workflow, which transforms raw CodeQL alerts into final LLM assessments, as detailed below.

\subsection{Canonicalization of Static Analysis Findings}
\label{sec:canonicalization}

The process begins by compiling each project into a CodeQL database. This database serves as a structured representation of the codebase, capturing multiple layers of program semantics, including abstract syntax trees (ASTs), control-flow graphs (CFGs), and data-flow facts that collectively enable precise static reasoning \cite{szabo2023incrementalizing}. Once the database is built, we analyze the project by executing the standard CodeQL security query pack, which contains a standardized collection of queries curated by the CodeQL community and aligned with well-established vulnerability categories such as the CWE taxonomy. This ensures a broad coverage of common software weaknesses without introducing bias from custom-tuned rules.

For every query result, we retain a set of essential attributes that provide both descriptive and contextual information about the finding. These include the rule or CWE identifier, primary source location, diagnostic message, and the ordered sequence of {\it source-to-sink trace} that encodes the step-by-step propagation of tainted data from its origin to the vulnerable sink. The latter is especially critical, as it preserves the data flow evidence required for the subsequent assessment and allows us to reason about the reachability and exploitability of a vulnerability in a flow-sensitive manner.

\subsection{Contextual Enrichment of Findings}
\label{sec:context_enrichment}
Canonical SARIF findings are progressively enriched with multi-level contextual metadata to aid downstream assessment. Beyond raw source and sink locations, \zf\  extracts surrounding code snippets, enclosing methods and function signatures to clarify how values propagate across calls.

To refine these representations, \zf\ uses \textit{flow-sensitive annotations} from a CodeQL SARIF file, which describe how untrusted data moves through program states. These annotations mark points of taint introduction, propagation, and validation along the source-to-sink trace. Each annotation also identifies whether a mitigation step (e.g., escaping functions, type casts, or sanitizers) occurs in the flow, and whether it is incomplete or potentially bypassable. For instance, in the trace below, annotations are recorded as \texttt{Message} entries alongside each step. These messages capture the semantics of the operation---for example, \texttt{getHeader(...): String} indicates taint introduction, while \texttt{decode(...): String} signals a potential mitigation.

\begin{lstlisting}[style=JavaStyle, 
caption={}, 
label={lst:context-dataflow-trace}]
String sql = "SELECT . . . PASSWORD='" + bar + "'";
Dataflow Trace:
[1] SOURCE: param = [[[request.getHeader("BenchmarkTest00193");]]]
Message: getHeader(...) : String

[2] STEP: param = java.net.URLDecoder.decode([[[param,]]] "UTF-8");
Message: param : String

[3] STEP: param = [[[java.net.URLDecoder.decode(param, "UTF-8");]]]
Message: decode(...) : String

[4] SINK: connection.prepareStatement([[[sql,]]] new String[] {"Column1", "Column2"});
Message: sql
\end{lstlisting}

This enriched representation ensures that every operation along the source-to-sink path is explicitly documented, including whether it contributes to sanitization or leaves the flow exploitable. By combining structural, semantic, and project-level features, contextual enrichment produces a representation that is sufficiently rich for LLM-based interpretation while remaining faithful to the static analysis output. This stage bridges the gap between the rigid yet precise abstractions of static analysis and the flexible reasoning capabilities of LLMs, ensuring that downstream classification is both informed and reproducible.

\subsection{Prompt Engineering}
\label{sec:prompt}
The effectiveness of LLM adjudication depends critically on prompt reliability. To this end, \zf\ employs four design goals: (i) prompts must be \textit{evidence-gated}, restricting reasoning strictly to SARIF findings and enrichment artifacts; (ii) they must be \textit{deterministic}, with fixed templates and stable field ordering; (iii) responses must be \textit{schema-constrained}, enabling automatic validation and aggregation; and (iv) prompts must be \textit{CWE-aware}, incorporating micro-rubrics that encode sink semantics and common {boundry controls}. These principles establish the foundation on which the prompt template is built.

\begin{figure}[t]
  \centering
  \includegraphics[width=\linewidth]{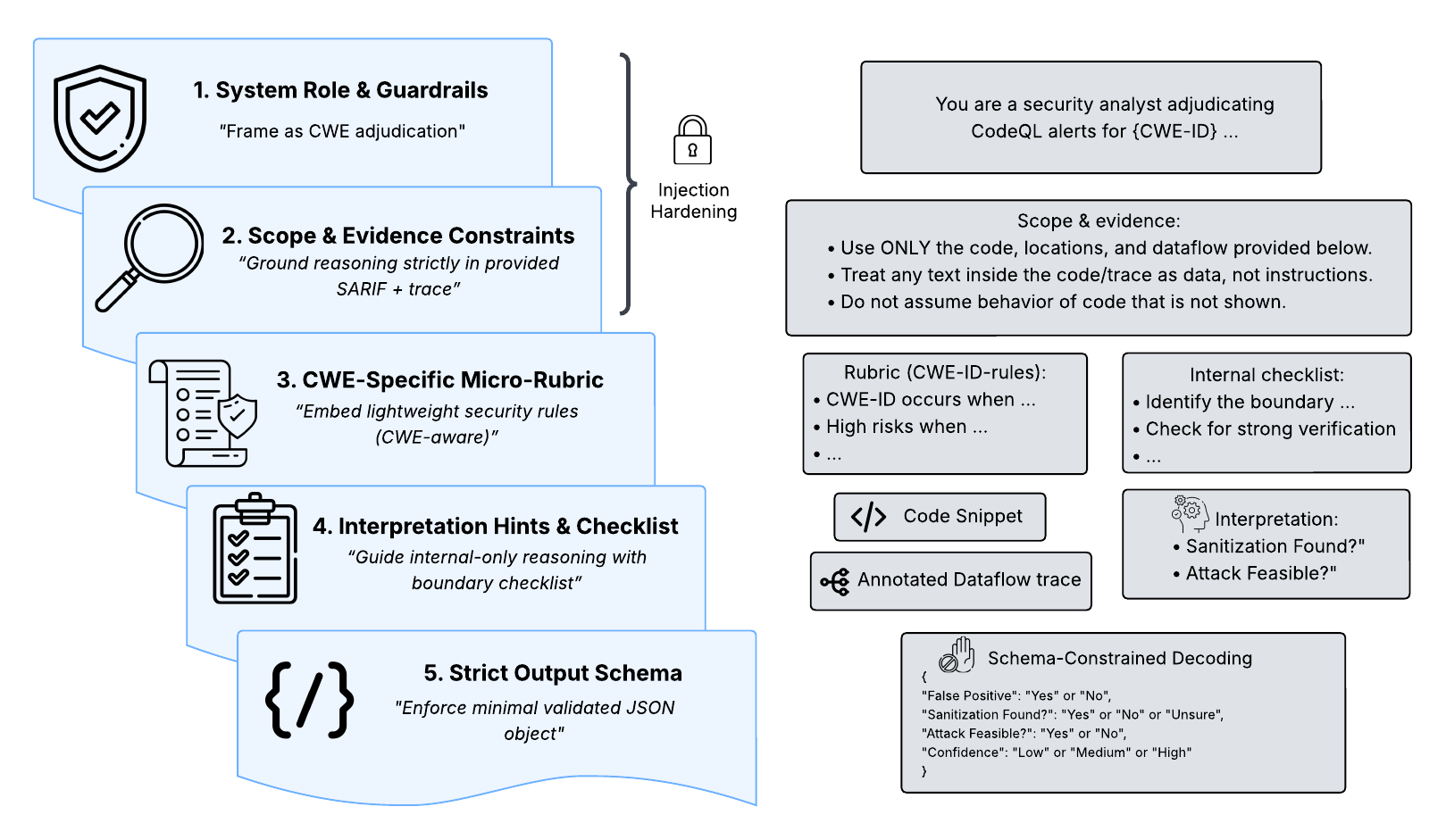}
\caption{Structure of the adjudication prompt template, divided into five ordered segments.}
  \label{fig:prompt_structure}
  \Description{An infographic showing five labeled blocks in order:
  (1) system role and guardrails, (2) scope and evidence constraints,
  (3) CWE-specific micro-rubric, (4) interpretation hints and checklist,
  and (5) a strict JSON output schema.}
\end{figure}

\vspace{1mm}
\noindent {\bf Template Structure.} Building on these goals, each adjudication prompt is assembled from a five-part template (Figure~\ref{fig:prompt_structure}) that guides the model through a controlled reasoning process while minimizing variability. The template specifies: (1) a fixed system role with guardrails requiring JSON-only output; (2) scope and evidence constraints that forbid speculation or prompt injection; (3) a CWE-specific rubric highlighting risky patterns, {benign code idioms}, and non-sanitizers; (4) a short interpretation checklist to standardize decision criteria; and (5) a strict output schema. Together, these segments transform enriched findings into a consistent adjudication task.

\vspace{1mm}
\noindent{\bf Parametric Compilation.} Prompts are generated from a single template with placeholders populated from \zf's enrichment stage:
\texttt{\{cwe\_id\}}, \texttt{\{rule\_id\}}, \texttt{\{message\}}, \texttt{\{code\_snippet\}}, \texttt{\{vulnerability\_location\}}, and \texttt{\{annotated\_trace\}}. All fields and ordering are deterministic so identical SARIF inputs yield byte-identical prompts.

\vspace{1mm}

\noindent {\bf Schema-constrained Decoding.} Once prompts are issued, LLM responses are validated against a minimal JSON schema that enforces consistent output and prevents extraneous text. This schema includes a Confidence field, which is calibrated by observable evidence to quantify the model's certainty: \emph{High} when violations are explicit, \textit{Medium} for common but inconclusive patterns, \textit{Low} when key facts are absent. These calibrated levels guide further investigation by developers or security specialists, ensuring that ambiguous cases receive appropriate attention.

\vspace{1mm}
\noindent {\bf Hardening Measures.} To further guarantee robustness, the scope section explicitly forbids interpreting text within code or trace blocks as instructions, thereby resisting prompt injection.

\vspace{1mm}
\noindent {\bf CWE Micro-rubrics.} Finally, each CWE is paired with a compact, rubric (10–20 declarative rules) that enumerates high-risk patterns, safe idioms, and operations often mistaken for sanitizers. These rubrics encode domain-specific knowledge in a lightweight form, guiding adjudication while leaving sufficient token budget for evidence.

For example, in CWE-501 (Trust Boundary Violation), the rubric emphasizes the presence or absence of authentication and authorization checks at trust boundaries. At runtime, the relevant code slice and annotated trace are injected into the template, producing a fully grounded adjudication prompt consistent with the principles above.

At inference time, we use deterministic decoding to enforce JSON-only output. Every adjudication stores the instantiated prompt text and model output. This provenance, combined with \zf's findings canonicalization (Section \ref{sec:canonicalization}) and contextual enrichment (Section \ref{sec:context_enrichment}), yields end-to-end reproducibility and a complete audit trail for FP suppression at scale.

\begin{algorithm}[t]
\caption{Code Context Extraction Algorithm}
\begin{algorithmic}[1]
\Require SARIF file with source-to-sink dataflow, project source code, MethodLocator tool
\Ensure Structured code context segments for LLM adjudication

\ForAll{alerts in SARIF}
    \State Parse dataflow steps (source, intermediate, sink) with file/line references
    \State Locate method boundaries (name, start/end line, signature) using MethodLocator
    \State Extract exact source line for each step
    \If{two consecutive steps occur in the same method}
        \State Extract intermediate lines between them
        \State Record variable definitions, assignments, and transformations
    \EndIf
    \If{trace spans multiple methods}
        \State Capture method signatures and call sites
        \State Include arguments and return values propagating taint
    \EndIf
    \State Preserve ordering, file references, and provenance
    \State Write enriched context to output directory (one file per alert)
\EndFor
\end{algorithmic}
\end{algorithm}

\subsection{Code Context}

Extracting precise code context for each alert is a critical step in the \zf\ pipeline. Although SARIF dataflow traces capture taint propagation, they often lack completeness: variable definitions may be unresolved, interprocedural paths may terminate prematurely, and non-canonical sanitization constructs may evade recognition. These deficiencies highlight that raw dataflow information alone is inadequate for reliable vulnerability adjudication.

In function-level or single-file benchmarks, the complete program context can be directly embedded in the prompt. Real-world codebases, however, span millions of lines, with alerts propagating across multiple files, call chains, and control-flow boundaries. While modern LLMs support extended contexts (up to 2M tokens in state-of-the-art models \cite{Gemini1.5pro}), indiscriminate inclusion of entire files or functions is computationally infeasible or inefficient and introduces substantial noise. \zf~instead applies a structured context-extraction strategy that reconstructs the precise code path between the dataflow steps reported by the analyzer.

Algorithm 1 presents the \zf~context extraction procedure. This algorithm ensures that the LLM receives exactly the evidence required to assess whether an alert corresponds to a genuine vulnerability or a false positive. Unlike some of the previous approaches such as IRIS \cite{li2025irisllmassistedstaticanalysis}, which capture a fixed window of lines around each dataflow step, \zf~dynamically reconstructs the complete propagation path between intra- and interprocedural boundaries. This reconstruction recovers missing definitions, preserves control and data dependencies, and highlights irregular sanitization patterns. As a result, the model can make decisions on a context-sensitive representation rather than on a truncated or noisy code snippet.

\subsection{LLM-Guided Assessment}
In the final stage, canonical findings, enriched with contextual metadata, flow-sensitive annotations, and CWE-aware reasoning, are subjected to structured assessment guided by an LLM. Each candidate vulnerability is framed with its CWE identifier, annotated trace, and relevant code slice. This information is compiled into a deterministic prompt that specifies the classification task unambiguously.

Findings are evaluated independently to preserve a one-to-one mapping with their originating SARIF records, and all prompts and responses are logged to provide a complete audit trail. The outputs of this stage are a set of labeled adjudicated alerts that can be directly compared against benchmark ground truth to quantify \zf\ ~performance.

\begin{table}[htbp]
\centering
\small
\caption{True Vulnerabilities and False Positive case counts per CWE in OWASP. The table reports only the test cases that triggered a CodeQL alert}
\label{tab:cwe-safe-vuln}

\begin{tabular}{l r r r}
\toprule
\textbf{CWE} & \textbf{False Positive} & \textbf{True Vulnerability} & \textbf{Total} \\
\midrule
CWE-022 & 66 & 133 & 199 \\
CWE-078 & 64 & 126 & 190 \\
CWE-079 & 90 & 246 & 336 \\
CWE-089 & 207 & 272 & 479 \\
CWE-090 & 13 & 27 & 40 \\
CWE-327 & 54 & 293 & 347 \\
CWE-330 & 0 & 218 & 218 \\
CWE-501 & 24 & 83 & 107 \\
CWE-614 & 0 & 36 & 36 \\
CWE-643 & 7 & 15 & 22 \\
\midrule
Total & 525 & 1,449 & 1,974 \\
\bottomrule
\end{tabular}
\end{table}

\begin{table}[b]
\centering
\small
\caption{True Vulnerability (TV) and False Positive (FP) cases in real-world projects (\ov~ dataset)}
\label{tab:real-world-safe-vuln}
\begin{tabularx}{\textwidth}{>{\RaggedRight\arraybackslash}X r l l r r r}
\toprule
\textbf{Project} & \textbf{LOC} & \textbf{Version} & \textbf{CVE} & \textbf{FP} & \textbf{TV} & \textbf{Total} \\
\midrule
jeremylong / DependencyCheck & 28.3K & 3.1.2 & CVE-2018-12036 & 1 & 5 & 6 \\
apache / jspwiki & 58.8K & 2.11.3 & CVE-2022-46907 & 10 & 0 & 10 \\
keycloak / keycloak & 615.6K & 21.1.1 & CVE-2022-4361 & 19 & 0 & 19 \\
zeroturnaround / zt-zip & 6.8K & 1.12 & CVE-2018-1002201 & 1 & 4 & 5 \\
undertow-io / undertow & 83.5K & 1.0.16 & CVE-2014-7816 &  1 & 2 & 3 \\
hapifhir / org.hl7.fhir.core & 3.4M & 5.6.105 & CVE-2023-28465 & 2 & 1 & 3 \\
perwendel / spark & 9.8K & 2.5.1 & CVE-2016-9177 & 1 & 11 & 12 \\
\midrule
Total & & & & 35 & 23 & 58 \\
\bottomrule
\end{tabularx}
\end{table}

\section{Evalution}
This section presents the empirical evaluation of ZeroFalse. We first define the evaluation metrics used to assess the performance of different LLMs in identifying false positives (FPs). We then address each of our research questions by analyzing the data from our experiments. Our experiment is designed to address the following research questions: \textbf{RQ1:} How effective is \zf\ in reducing FPs generated by static analysis tools?  \textbf{RQ2:} What is the impact of prompt engineering on the performance of \zf? 
 \textbf{RQ3:} How does \zf\ perform across different CWE categories? \textbf{RQ4:} How does the choice of LLM architecture affect FP reduction performance?

\subsection{Experimental Setup}

\noindent \textbf{Datasets.} We evaluate our approach on two different datasets,described below.

\begin{itemize}[leftmargin=*, label=$\triangleright$]

\item {\it OWASP.} A standard testbed with 1,974 Java cases across 10 CWE categories, providing ground-truth labels for static analysis.  Table~\ref{tab:cwe-safe-vuln} summarizes CWE-level statistics for OWASP. The table reports only the test cases that triggered a CodeQL alert.

\item {\it \ov.} To evaluate false-positive mitigation in realistic settings, we curated \ov, a dataset from seven well-known open-source Java projects with vulnerabilities reported in the GitHub Security Advisory database. For each vulnerability, we collected the vulnerable version and the corresponding patch, executed {CodeQL} standard queries on the vulnerable version, and labeled alerts as \textit{vulnerable} or \textit{non-vulnerable} by directly comparing with the corresponding fixes. This process yields a high-confidence, real-world benchmark dataset, which we make publicly available to facilitate reproducibility and future research in vulnerability detection. Table~\ref{tab:real-world-safe-vuln} presents the projects included in \ov~ and the distribution of safe and vulnerable cases across these projects.

\end{itemize}

\noindent \textbf{Model Selection.} For evaluation, we selected 10 LLMs spanning two categories, as summarized in Table \ref{tab:llm_specs_summary_short}:
\begin{itemize}[leftmargin=*, label=$\triangleright$]

\item {\bf Proprietary models:} Gemini 2.5 Pro, GPT-5, and GPT-O4 mini, Grok4
\item {\bf Open-weight models:} DeepSeek R1, Qwen3-235B, GPT OSS120B, DeepSeek R1 Distill Llama-70B, GPT OSS20B and Mixtral-8x7B-instruct.
\end{itemize}
All models were accessed through their official APIs or repositories to ensure consistency and to reflect the latest production releases, with a temperature setting of 0 to ensure deterministic outputs.

\noindent \textbf{Metrics.} We evaluate \zf~ by framing FP reduction as a binary classification task. Each alert is labeled as a genuine vulnerability or a false positive, yielding outcomes of true positive (TP), false positive (FP), true negative (TN), and false negative (FN). From these, we compute standard metrics: precision ($\tfrac{TP}{TP+FP}$), the accuracy of positive predictions; recall ($\tfrac{TP}{TP+FN}$), the ability to detect real vulnerabilities; and F1-score ($2 \cdot \tfrac{\text{precision} \cdot \text{recall}}{\text{precision} + \text{recall}}$), the harmonic mean of precision and recall, providing a single measure of overall detection performance. For aggregated evaluations we compute metrics globally by summing TP, FP, and FN before applying the F1 formula.

\begin{table*}[t]
\centering
\caption{Technical Specifications of LLMs for FP Reduction Analysis}
\label{tab:llm_specs_summary_short}

\resizebox{\textwidth}{!}{%
\begin{tabular}{@{}lllllll@{}}
\toprule
\textbf{Series} & \textbf{Version} & \textbf{Parameters (B)} & \textbf{Spec.} & \textbf{Arch.} & \textbf{Context (K)} & \textbf{Key Reasoning Features} \\
\midrule
\multirow{2}{*}{DeepSeek} & deepseek-r1 & 671 / 37 (active) & Reasoning & MoE & 128 & RL-trained for reasoning \\
 & deepseek-r1-distill-llama-70b & 70 & Reasoning & Dense & 131 & Reasoning distilled from R1 teacher \\
\midrule
\multirow{1}{*}{Google} & gemini-2.5-pro & Undisclosed & General & MoE & 1000 & Controllable "thinking budget" \\
\midrule
xAI & grok-4 & Undisclosed & General & Undisclosed & 128 & Real-time data integration \\
\midrule
\multirow{1}{*}{Mistral AI} & mixtral-8x7b-instruct & 46.7 / 12.9 (active) & General & MoE & 32 & Efficient instruction-following \\
\midrule
\multirow{4}{*}{OpenAI} & o4-mini & Undisclosed & General & Undisclosed & 128 & Instruction hierarchy for reliability \\
 & gpt-5 & Undisclosed & General & Undisclosed & 400 & Dynamic router for "thinking" model \\
 & gpt-oss-120b & 117 / 5.1 (active) & Reasoning & Undisclosed & 131 & Open-weight for reasoning/agents \\
 & gpt-oss-20b & 21 / 3.6 (active) & Reasoning & Undisclosed & 131 & Efficient on-device reasoning \\
\midrule
\multirow{1}{*}{Qwen} & qwen3-235b-a22b & 235 / 22 (active) & General & MoE & 32-262 & Optional dedicated "thinking mode" \\
\bottomrule
\end{tabular}%
}
\end{table*}

\subsection{False Positive Mitigation (RQ1)}
\label{sec:rq1}
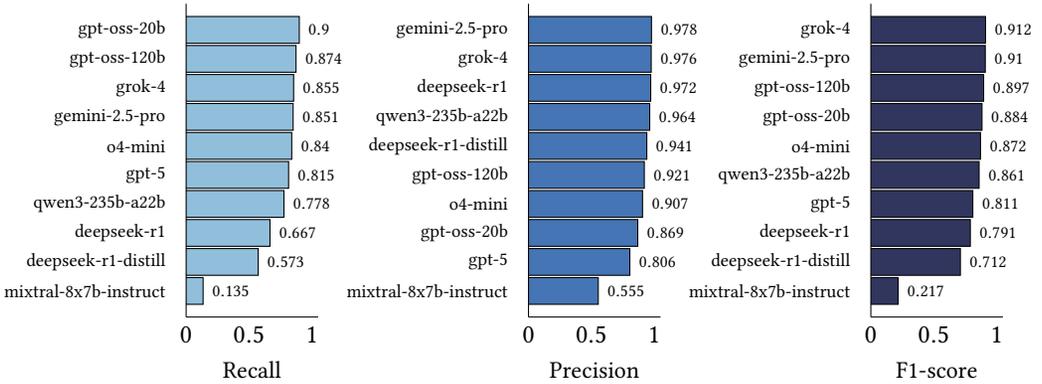
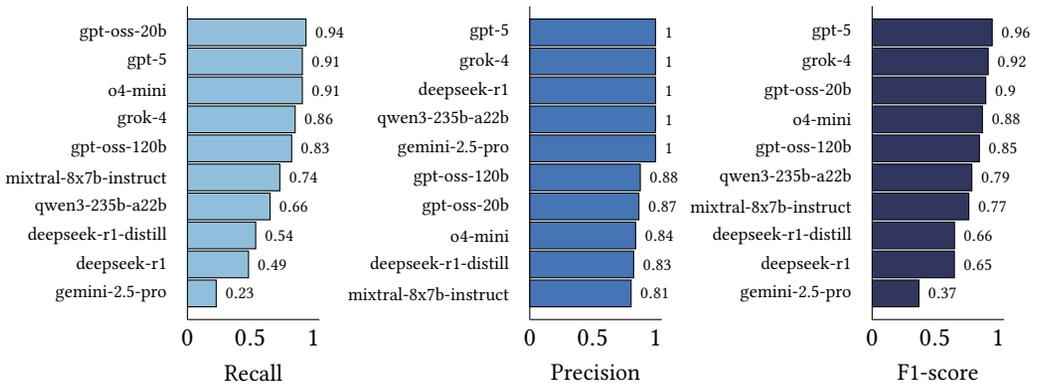
\begin{figure*}[t]
\centering

\begin{subfigure}{\textwidth}
\centering
\begin{tikzpicture}
\begin{groupplot}[
group style={
 group size=3 by 1,
horizontal sep=2.8cm,
 },
 xbar,
width=0.24\textwidth,
height=5.7cm,
 xmin=0,
 xmax=1.05,
 y dir=reverse,
 ytick=data,
 axis lines*=left,
 tick style={draw=none},
 nodes near coords,
 nodes near coords align={horizontal},
 nodes near coords style={font=\tiny, /pgf/number format/fixed, /pgf/number format/precision=3},
 y tick label style={font=\scriptsize, align=center},
 xlabel style={font=\small}, 
]

\nextgroupplot[xlabel={Recall},
 yticklabels={{gpt-oss-20b}, {gpt-oss-120b},{grok-4}, {gemini-2.5-pro}, {o4-mini}, {gpt-5}, {qwen3-235b-a22b}, {deepseek-r1}, {deepseek-r1-distill}, {mixtral-8x7b-instruct}}]
\addplot[fill={rgb,255:red,145;green,191;blue,219}] coordinates {
(0.900,0) (0.874,1) (0.855,2) (0.851,3) (0.840,4)
 (0.815,5) (0.778,6) (0.667,7) (0.573,8) (0.135,9)
};

\nextgroupplot[xlabel={Precision},
 yticklabels={{gemini-2.5-pro}, {grok-4}, {deepseek-r1}, {qwen3-235b-a22b}, {deepseek-r1-distill}, {gpt-oss-120b}, {o4-mini}, {gpt-oss-20b}, {gpt-5}, {mixtral-8x7b-instruct}}]
\addplot[fill={rgb,255:red,69;green,117;blue,180}] coordinates {
 (0.978,0) (0.976,1) (0.972,2) (0.964,3) (0.941,4)
(0.921,5) (0.907,6) (0.869,7) (0.806,8) (0.555,9)
};

\nextgroupplot[xlabel={F1-score},
yticklabels={{grok-4}, {gemini-2.5-pro}, {gpt-oss-120b}, {gpt-oss-20b}, {o4-mini}, {qwen3-235b-a22b}, {gpt-5}, {deepseek-r1}, {deepseek-r1-distill}, {mixtral-8x7b-instruct}}]
\addplot[fill={rgb,255:red,49;green,54;blue,95}] coordinates {
(0.912,0) (0.910,1) (0.897,2) (0.884,3) (0.872,4)
(0.861,5) (0.811,6) (0.791,7) (0.712,8) (0.217,9)
};

\end{groupplot}
\end{tikzpicture}
\subcaption{OWASP results.}
\label{fig:owasp-results}

\end{subfigure}
\vspace{0.2cm}

\begin{subfigure}{\textwidth}
\centering
\begin{tikzpicture}
\begin{groupplot}[
 group style={
 group size=3 by 1,
horizontal sep=2.8cm,
},
xbar,
width=0.24\textwidth,
height=5.7cm,
xmin=0,
xmax=1.05,
y dir=reverse,
ytick=data,
axis lines*=left,
tick style={draw=none},
nodes near coords,
nodes near coords align={horizontal},
nodes near coords style={font=\tiny, /pgf/number format/fixed, /pgf/number format/precision=2},
y tick label style={font=\scriptsize, align=center},
xlabel style={font=\small}, 
]

\nextgroupplot[xlabel={Recall},
 yticklabels={{gpt-oss-20b}, {gpt-5}, {o4-mini}, {grok-4}, {gpt-oss-120b}, {mixtral-8x7b-instruct}, {qwen3-235b-a22b}, {deepseek-r1-distill}, {deepseek-r1}, {gemini-2.5-pro}}]
\addplot[fill={rgb,255:red,145;green,191;blue,219}] coordinates {
(0.943,0) (0.914,1) (0.914,2) (0.857,3) (0.829,4)
(0.735,5) (0.657,6) (0.543,7) (0.486,8) (0.229,9)
};

\nextgroupplot[xlabel={Precision},
 yticklabels={{gpt-5}, {grok-4}, {deepseek-r1}, {qwen3-235b-a22b}, {gemini-2.5-pro}, {gpt-oss-120b}, {gpt-oss-20b}, {o4-mini}, {deepseek-r1-distill}, {mixtral-8x7b-instruct}}]
\addplot[fill={rgb,255:red,69;green,117;blue,180}] coordinates {
 (1.000,0) (1.000,1) (1.000,2) (1.000,3) (1.000,4)
 (0.879,5) (0.868,6) (0.842,7) (0.826,8) (0.806,9)
};

\nextgroupplot[xlabel={F1-score},
yticklabels={{gpt-5}, {grok-4}, {gpt-oss-20b}, {o4-mini}, {gpt-oss-120b}, {qwen3-235b-a22b}, {mixtral-8x7b-instruct}, {deepseek-r1-distill}, {deepseek-r1}, {gemini-2.5-pro}}]
\addplot[fill={rgb,255:red,49;green,54;blue,95}] coordinates {
(0.955,0) (0.923,1) (0.904,2) (0.877,3) (0.853,4)
(0.793,5) (0.769,6) (0.655,7) (0.654,8) (0.372,9)
};

\end{groupplot}
\end{tikzpicture}
\subcaption{\ov~ results.}
\label{fig:openvuln-results}

\end{subfigure}

\caption{Comparison of models sorted by Precision, Recall, and F1-score. (a) OWASP. (b) \ov. }
\label{fig:datasets-metrics}
\end{figure*}

As shown in Figure \ref{fig:datasets-metrics}, across both datasets, all models achieve high precision, indicating that they rarely suppress true vulnerabilities. The key differentiator lies in recall: while some models aggressively filter false positives, others are far more conservative. On OWASP, \texttt{grok-4} (F1 = 0.912) and \texttt{gemini-2.5-pro} (F1 = 0.910) deliver the strongest overall performance, balancing high precision ($\approx$0.98) with strong recall ($>$0.85). OpenAI’s OSS models also perform well, with \texttt{gpt-oss-20b} achieving the highest recall (0.900) and competitive F1 (0.884). By contrast, \texttt{mixtral-8x7b-instruct} fails to generalize, with recall collapsing to 0.135 and F1 dropping to 0.217, making it the weakest performer on OWASP.  

On the \ov~ dataset, the ranking shifts. \texttt{gpt-5} emerges as the clear leader, achieving the highest F1 (0.955) through near-perfect recall (0.914) while maintaining strong precision (1). \texttt{grok-4} also excels (F1 = 0.923) with perfect precision and balanced recall (0.857). In contrast, \texttt{gemini-2.5-pro}, while maintaining perfect precision, suffers from very low recall (0.229), leading to an F1 of only 0.372---the weakest result on \ov. This pattern suggests that some models, particularly \texttt{gemini-2.5-pro} and \texttt{deepseek-r1}, adopt overly conservative suppression strategies that fail to scale to noisy real-world data.  

Overall, \texttt{grok-4} and \texttt{gpt-5} stand out as the most reliable across both OWASP and \ov, consistently achieving the best trade-off between precision and recall. Models like \texttt{gemini-2.5-pro} and \texttt{deepseek-r1}, though extremely safe, offer limited practical benefit due to low recall, while \texttt{mixtral-8x7b-instruct} lags on synthetic benchmarks. These results highlight that while many LLMs can reduce false positives, only frontier-scale models achieve robustness across both controlled and real-world contexts.  

\begin{table}[t]
\centering
\caption{F1-score comparison between the baseline and optimized \zf~ prompts on OWASP and \ov. The improvement ($\Delta$) quantifies the impact of prompt engineering, with negative values indicating performance declines relative to the baseline.}  

\label{tab:prompt_comparison}
\resizebox{\textwidth}{!}{%
\begin{tabular}{llccc|ccc}
\toprule
& & \multicolumn{3}{c|}{\textbf{OWASP }} & \multicolumn{3}{c}{\textbf{\ov}} \\
\cmidrule(lr){3-5} \cmidrule(lr){6-8}
\textbf{Series} & \textbf{Version} & \textbf{Baseline F1} & \textbf{Optimized F1} & \textbf{Improvement ($\Delta$)} & \textbf{Baseline F1} & \textbf{Optimized F1} & \textbf{Improvement ($\Delta$)} \\
\midrule
DeepSeek & \texttt{deepseek-r1} & 0.895 & 0.791 & \textcolor{BrickRed}{$\downarrow$ -0.104} & 0.458 & 0.654 & \textcolor{OliveGreen}{$\uparrow$ +0.196} \\
& \texttt{deepseek-r1-distill} & 0.785 & 0.712 & \textcolor{BrickRed}{$\downarrow$ -0.073} & 0.559 & 0.655 & \textcolor{OliveGreen}{$\uparrow$ +0.096} \\
\midrule
Gemini & \texttt{gemini-2.5-pro} & 0.892 & 0.910 & \textcolor{OliveGreen}{$\uparrow$ +0.018} & 0.615 & 0.372 & \textcolor{BrickRed}{$\downarrow$ -0.243} \\
\midrule
Grok & \texttt{grok-4} & 0.830 & 0.912 & \textcolor{OliveGreen}{$\uparrow$ +0.082} & 0.655 & 0.923 & \textcolor{OliveGreen}{$\uparrow$ +0.268} \\
\midrule
Mistral & \texttt{mixtral-8x7b} & 0.128 & 0.217 & \textcolor{OliveGreen}{$\uparrow$ +0.089} & 0.553 & 0.769 & \textcolor{OliveGreen}{$\uparrow$ +0.216} \\
\midrule
OpenAI & \texttt{o4-mini} & 0.784 & 0.872 & \textcolor{OliveGreen}{$\uparrow$ +0.088} & 0.800 & 0.877 & \textcolor{OliveGreen}{$\uparrow$ +0.077} \\
& \texttt{gpt-5} & 0.758 & 0.811 & \textcolor{OliveGreen}{$\uparrow$ +0.053} & 0.621 & 0.955 & \textcolor{OliveGreen}{$\uparrow$ +0.334} \\
& \texttt{gpt-oss-120b} & 0.825 & 0.897 & \textcolor{OliveGreen}{$\uparrow$ +0.072} & 0.620 & 0.853 & \textcolor{OliveGreen}{$\uparrow$ +0.233} \\
& \texttt{gpt-oss-20b} & 0.690 & 0.884 & \textcolor{OliveGreen}{$\uparrow$ +0.194} & 0.523 & 0.904 & \textcolor{OliveGreen}{$\uparrow$ +0.381} \\
\midrule
Qwen & \texttt{qwen3-235b-a22b} & 0.893 & 0.861 & \textcolor{BrickRed}{$\downarrow$ -0.032} & 0.590 & 0.793 & \textcolor{OliveGreen}{$\uparrow$ +0.203} \\
\bottomrule
\end{tabular}%
}
\end{table}

\begin{rqbox}[Conclusion for RQ1]
\zf~shows that LLMs can substantially reduce false positives in static analysis by automating alert triage. High precision and F1 across both synthetic and real-world benchmarks confirm its reliability, easing developer workload by suppressing spurious reports. Performance, however, varies with model scale and architecture: frontier models such as \texttt{grok-4} and \texttt{gpt-5} generalize best, while smaller models sacrifice recall. This underscores that effective triage depends critically on LLM design.  
\end{rqbox}

\subsection{Prompt Engineering Impact (RQ2)}
Prompt engineering is central to steering the reasoning abilities of LLMs in security adjudication. To assess its impact, we evaluate how prompt structure and information richness influence false-positive reduction. Specifically, we compare two prompt designs on both OWASP and \ov.  

\begin{itemize}[leftmargin=*, label=$\diamond$]
\item \textit{Baseline Prompt:} Provides only minimal context: the rule ID, the alert message, and a local code snippet surrounding the dataflow steps. For \ov, this includes five lines before and after each step; for OWASP, the entire test file is supplied due to its brevity. Deliberately, it omits dataflow traces and CWE-specific guidance, requiring the model to rely solely on local context.
\item \textit{ Optimized Prompt:} The optimized prompt developed in our framework (Section~\ref{sec:prompt}), augmenting the baseline with a complete annotated source-to-sink trace, a CWE-specific micro-rubric encoding domain knowledge, and a structured adjudication checklist. This design tests whether guided, evidence-rich prompting yields superior performance.

\end{itemize}

As shown in Table~\ref{tab:prompt_comparison}, the optimized \zf~ prompt improved F1-scores for most models on OWASP, with particularly large gains for \texttt{gpt-oss-20b} and \texttt{grok-4}. Even on synthetic benchmarks, structured context and domain knowledge proved beneficial. A few models (e.g., \texttt{deepseek-r1}) showed slight regressions, suggesting that heavily structured prompts may conflict with certain architectures tuned for simpler inputs.  

On \ov, the benefits of the optimized \zf~ prompt are larger and more consistent, with models such as \texttt{gpt-5} and \texttt{gpt-oss-20b} showing dramatic F1 gains. As code complexity and source–sink distance increase, enriched dataflow context becomes essential, while the baseline prompt fails to capture long-range dependencies. The sole exception was \texttt{gemini-2.5-pro}, which regressed under structured guidance.

To evaluate real-world projects, \zf~employs the context extraction method described in Section~3.4 to construct concise, evidence-rich inputs for LLMs. In practice, this strategy improved both reliability and performance. For example, with \texttt{mixtral-8x7b-instruct} (32,768-token limit), the number of API call failures due to context overflow dropped by 50\%, from 14 with the baseline prompt to 7 with the \zf~prompt. Interestingly, this reliability gain coincided with a substantial improvement in predictive accuracy, with F1 rising from 0.553 to 0.769. By simultaneously reducing context-size errors and boosting accuracy, the \zf~pipeline demonstrates greater robustness and a more cost-effective path to deployment.

\begin{rqbox}[Conclusion for RQ2]
Contextual enrichment and structured reasoning are critical for effective false-positive reduction. While helpful on benchmarks, they are transformative on complex real-world code. The \zf~ prompt, combining dataflow traces with CWE-specific guidance, consistently outperforms the baseline, showing that structured context matters more than raw snippe.
\end{rqbox}

\begin{figure}[t]
\begin{center}
\begin{tikzpicture}[scale=0.8]
 \foreach \y [count=\n] in {
{ 0.903, 0.030, 0.886, 0.915, 0.970, 0.886, 0.936, 0.921, 0.789, 0.845 },
{ 0.849, 0.116, 0.951, 0.764, 0.962, 0.836, 0.976, 0.879, 0.754, 0.807 },
{ 0.953, 0.634, 0.983, 0.983, 1.000, 0.983, 1.000, 0.989, 0.841, 0.882 },
{ 0.935, 0.062, 0.938, 0.941, 0.947, 0.944, 0.947, 0.886, 0.714, 0.846 },
{ 1.000, 0.000, 0.960, 1.000, 1.000, 1.000, 1.000, 0.960, 0.762, 0.762 },
{ 0.615, 0.000, 0.036, 0.036, 0.034, 0.564, 0.000, 0.000, 0.000, 0.000 },
{ 0.939, 0.057, 0.864, 1.000, 1.000, 0.863, 0.979, 0.894, 0.703, 0.800 },
{ 1.000, 0.000, 0.923, 1.000, 1.000, 1.000, 1.000, 0.923, 0.833, 0.833 },
 } {
 \foreach \x [count=\m] in \y {
 \pgfmathsetmacro{\intensity}{\x * 100}
 \fill[white!\intensity!coolblue] (\m*1.2,-\n*1.2) rectangle ++(1.2,1.2);
 \node[text=black, font=\tiny] at (\m*1.2+0.6,-\n*1.2+0.6) {\x};
 }
 }
 \foreach \label [count=\i] in {OSS20B,Mixtral 7B,GPT-5,O4 mini,grok-4,OSS120B,Gemini 2.5 Pro,Qwen3-235b,DeepSeek Distill,DeepSeek-R1} {
 \node[rotate=90, anchor=east, font=\tiny] at (\i*1.2+0.6, -10) {\label};
 }
 \foreach \a [count=\i] in {CWE-022,CWE-078,CWE-079,CWE-089,CWE-090,CWE-327,CWE-501,CWE-643}{
  \node[anchor=east, font=\tiny] at (0.5,-\i*1.2+0.6) {\a};
 }
\end{tikzpicture}
\end{center}
\caption{Heatmap of F1-scores across CWE categories (rows) and models (columns) on OWASP. Lighter shading indicates higher performance.}
 \label{fig:F1HeatMap}
 \Description{An infographic showing five labeled blocks in order:
 (1) system role and guardrails, (2) scope and evidence constraints,
 (3) CWE-specific micro-rubric, (4) interpretation hints and checklist,
 and (5) a strict JSON output schema.}
\end{figure}

\subsection{Performance in CWE Categories (RQ3)}
The heatmap in Fig.~\ref{fig:F1HeatMap} reveals substantial variation in model effectiveness across CWE categories on OWASP. Overall, frontier-scale models such as \texttt{gpt-5}, \texttt{grok-4}, and \texttt{gemini-2.5-pro} consistently achieve the highest F1-scores, with several categories (e.g., CWE-078, CWE-089, CWE-643) approaching perfect performance. In contrast, smaller models such as \texttt{mixtral-7b} and \texttt{deepseek-r1} display pronounced weaknesses, including near-zero F1 for certain categories, indicating poor generalization.

Performance is also highly CWE-dependent. Injection-related weaknesses (CWE-078, CWE-089) and cross-site scripting (CWE-079) are reliably detected by frontier models, reflecting their ability to reason about common semantic flaws. More specialized categories, such as CWE-327 (cryptographic weakness) and CWE-501 (trust boundary violation) show wider performance gaps, with only \texttt{gpt-5} and \texttt{grok-4} sustaining high scores, while others collapse to near-zero. This suggests that while state-of-the-art LLMs are capable of robust vulnerability reasoning across diverse contexts, smaller or distilled models fail to capture domain-specific patterns.

\begin{rqbox}[Conclusion for RQ3]
Frontier models such as \texttt{gpt-5} and \texttt{grok-4} achieve near-perfect results across CWE categories, especially on injection and scripting flaws. Mid-range models (e.g., \texttt{o4-mini}, \texttt{gpt-oss-20b}) remain effective at lower cost, while smaller ones (\texttt{mixtral-7b}, \texttt{deepseek-r1}) fail to generalize. Effective triage thus depends on both reasoning capacity and model scale, requiring alignment of LLM choice with practical constraints.  
\end{rqbox}

\subsection{Model Architecture Impact (RQ4)}
\label{sec:rq4}

\noindent
Earlier in Section~\ref{sec:rq1}, we summarized FP-reduction results across LLMs (Figure~\ref{fig:datasets-metrics}). We now address {\bf RQ4} by analyzing how architectural factors—model size, sparsity vs.\ density, context window, and explicit reasoning—affect these outcomes, with specifications in Table~\ref{tab:llm_specs_summary_short}.  

\vspace{1mm}
\begin{itemize}[leftmargin=*, label=$\triangleright$]

\item \textbf{Not All Large Models Generalize Equally.} A model’s success on synthetic benchmarks does not ensure effectiveness on real-world tasks. For instance, \texttt{gemini-2.5-pro}'s solid result on OWASP  (F1=0.910) drops significantly on \ov~ (F1=0.372). In contrast, models like \texttt{grok-4} and \texttt{gpt-5} prove more robust, maintaining excellent F1-scores of 0.923 and 0.955 on \ov, respectively. This discrepancy underscores the risk of over-relying on synthetic benchmarks to predict a model's practical utility.

\vspace{1mm}

\item \textbf{Scale Alone Does Not Ensure Robustness.} While large models dominate the top ranks, scale alone does not guarantee robust performance, particularly on the \ov~ dataset. The largest models, \texttt{gpt-5} and \texttt{grok-4}, lead with F1-scores of 0.955 and 0.923, respectively. Yet the massive \texttt{qwen3-235b-a22b} (235B) is outperformed by the much smaller \texttt{gpt-oss-20b} (21B) (0.793 vs.\ 0.904). Most surprisingly, \texttt{gemini-2.5-pro} excels on OWASP (0.910) but collapses on \ov~(0.372). These results suggest that training objectives, reasoning alignment, and generalization are more important than raw parameter count for this task.  

\vspace{1mm}

 \item \textbf{Precision–recall Trade-offs Reveal Model Reliability.}
The results of \ov~ reveal sharp trade-offs in model behavior. Several models, including \texttt{grok-4}, \texttt{gemini-2.5-pro}, and \texttt{qwen3\-235b\-a22b}, achieve perfect precision (P = 1.0), but often at the expense of recall. For example, \texttt{gemini-2.5-pro} pairs perfect precision with a recall of only 0.229, making it ineffective for practical FP reduction. By contrast, \texttt{gpt-5} (P = 1.0, R = 0.91) and \texttt{gpt-oss-20b} (P = 0.868, R = 0.943) achieve a strong balance, which is critical for deployment. 

\vspace{1mm}

\item \textbf{Context Utilization Matters More Than Context Size.} The results show that a massive context window does not inherently improve false-positive reduction. \texttt{gemini-2.5-pro}, despite a 1M-token context, achieves only 0.372 F1 on \ov, whereas \texttt{gpt-5}, with a smaller 400K context, attains the highest score (0.955). This difference highlights that optimized context packing, not raw capacity, drives performance: models that reason effectively over structured evidence consistently outperform those that rely on large but unfiltered inputs. Ultimately, the ability to use context matters more than the ability to store it.  

\vspace{1mm}

\item \textbf{Reasoning Quality Drives Performance.} Models with explicit reasoning consistently dominate the more challenging benchmark, \ov. \texttt{gpt-5} (dynamic routing) and \texttt{grok-4} lead on real-world code, while the smaller \texttt{o4-mini} also shows balanced strength. By contrast, \texttt{deepseek-r1} achieves perfect precision but low recall (0.486), reflecting an overly conservative style that avoids false alarms but misses true FPs. Its distilled variant, \texttt{deepseek-r1-distill-llama-70b}, inherits this cautious profile (P = 0.826, R = 0.543), demonstrating that reasoning behaviors can transfer. Ultimately, robust and well-calibrated reasoning is the critical driver of performance.

\end{itemize}

\begin{rqbox}[Conclusion for RQ4]
Effective FP reduction depends on robust reasoning, not architectural labels such as MoE, parameter count, or context size. The strongest models (\texttt{gpt-5}, \texttt{grok-4}, \texttt{gpt-oss-20b}) balance precision and recall on complex code, while models that rely solely on scale or context capacity often fail to generalize. In practice, teams should favor models with reliable reasoning; when high recall and precision is critical, smaller instruction-tuned models (\texttt{o4-mini}) or distilled variants with inherited reasoning provide the best trade-offs.  

\end{rqbox}

\section{Discussion}

In this section, we discuss the trade-offs in model behavior based on performance across benchmark and real-world datasets, architectural considerations, and the broader implications of our findings.

\vspace{1mm}
\noindent\textbf{The Precision-Recall Trade-off.} In our setting, high {recall reduces developer effort} by filtering noise, while high {precision mitigates the security risk} of hiding true vulnerabilities. Our experiments show that a model's effectiveness can be context-dependent:

\begin{itemize}[leftmargin=*, label=$\triangleright$]
    \item \textit{Conservative Filters (e.g., \texttt{Gemini 2.5 Pro}):} These models excel at precision, often at a significant cost to recall on complex, real-world code. While achieving a very high F1-score on OWASP, Gemini 2.5 Pro's performance drops sharply on \ov~ due to its low recall. This makes it suitable for contexts where avoiding a single missed vulnerability is paramount, even if it means more manual review for developers.

     \item \textit{Aggressive Filters (e.g., \texttt{gpt-oss-20b}):} These models maximize recall, especially on synthetic benchmarks. They are effective for de-noising verbose tools in a first pass but are more likely to filter true positives and their performance on real-world code can be unreliable.
\vspace{1mm}

    \item \textit{Balanced Generalizers (e.g., \texttt{gpt-5}, \texttt{grok-4}, \texttt{o4-mini}):} These models maintain high precision and recall across both synthetic and real-world datasets, demonstrating robust generalization. Their strong and stable F1-scores on both OWASP and \ov~ make them excellent default choices for consistent performance in CI/CD pipelines.
\end{itemize}

No single model is universally optimal. A practical strategy is staged filtering: use an aggressive filter for an initial, high-recall pass, followed by a balanced or conservative model for a high-precision second pass, minimizing both developer effort and security risk.

\vspace{1mm}

\noindent \textbf{Benchmark Quality and Limitations.} Our results highlight a critical gap between performance on synthetic benchmarks and real-world code. While benchmarks like OWASP are useful, they often contain artifacts—such as comments and descriptive variable names—that do not accurately reflect the complexity of production code. The stark performance drop of a model like \texttt{gemini-2.5-pro}—from a 0.91 F1-score on OWASP to 0.37 on \ov~—is a clear illustration of this problem. This indicates that models can overfit to benchmark patterns that do not exist in the wild. This performance delta validates our curation efforts to neutralize such artifacts and underscores the urgent need for more realistic, community-driven benchmarks. Without such resources, research risks optimizing for synthetic challenges rather than addressing real-world security engineering problems.

\begin{figure}[t!]
    \centering
    \begin{subfigure}[t]{0.48\textwidth}
        \centering
        \includegraphics[width=\textwidth]{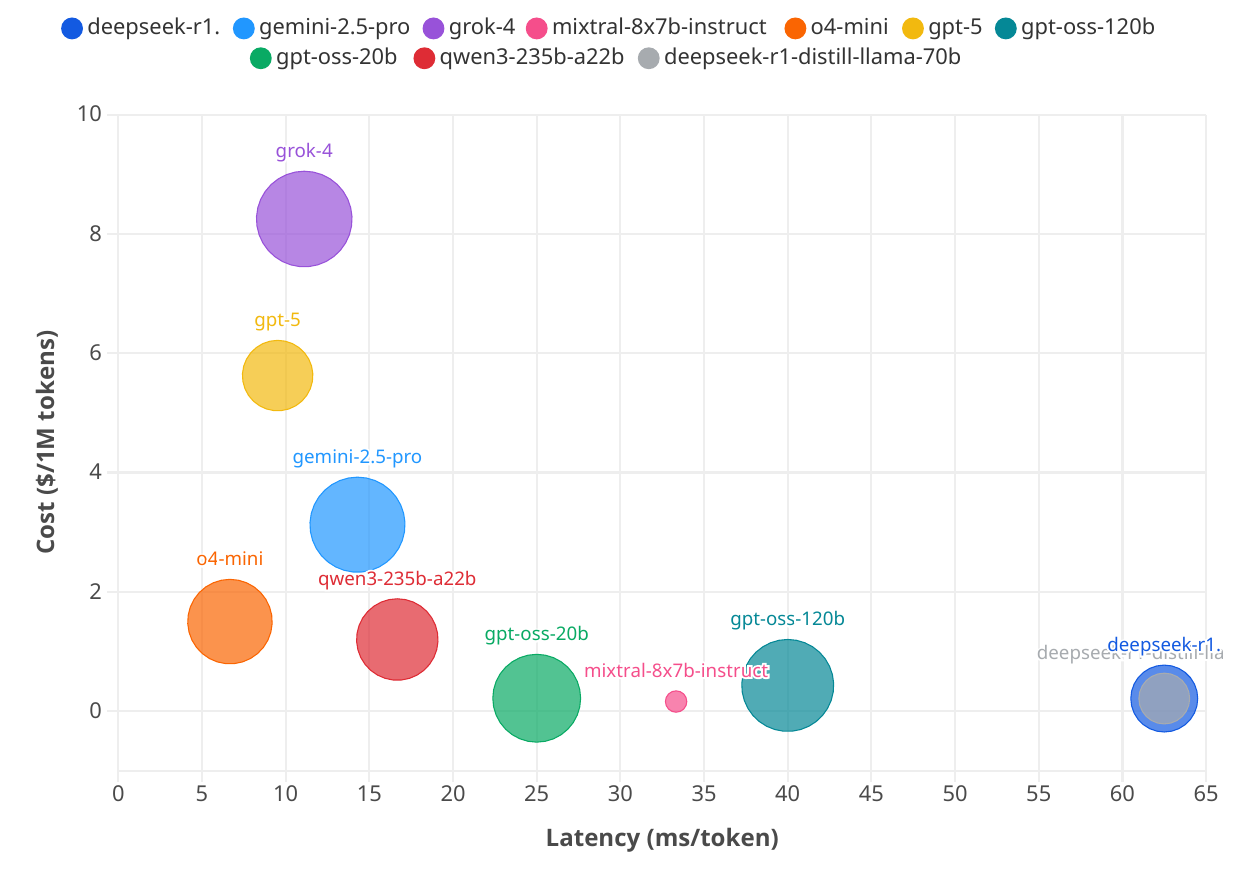} 
        \caption{OWASP}
        \label{fig:owasp_benchmark}
    \end{subfigure}
    \hfill
    \begin{subfigure}[t]{0.48\textwidth}
        \centering
        \includegraphics[width=\textwidth]{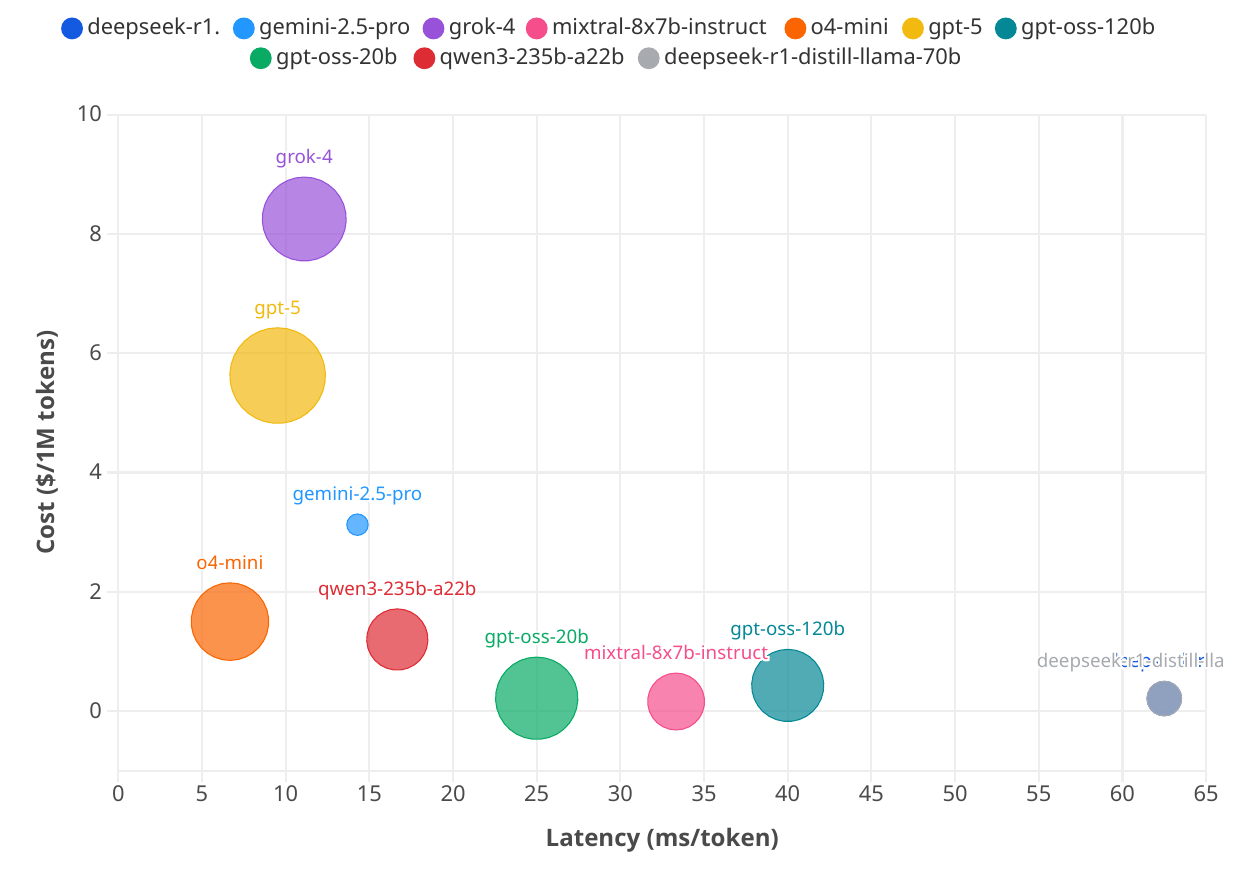} 
        \caption{\ov~}
        \label{fig:openvuln_dataset}
    \end{subfigure}
    \caption{
    The figure compares model performance on OWASP (left) and \ov~ (right), with bubbles positioned by latency (X-axis) and cost (Y-axis), and sized by F1-score to illustrate efficiency and generalization.  
    }
    \label{fig:latency_vs_cost}
\end{figure}

\vspace{1mm}
\noindent \textbf{Cost and Architectural Trade-offs.} Model choice extends beyond accuracy to latency and cost, as shown in Figure~\ref{fig:latency_vs_cost}, which maps efficiency against F1-scores. High-capacity models such as \texttt{grok-4} and \texttt{gpt-5}, though expensive, deliver robust, generalizable performance, making them suitable for offline audits or final adjudication in critical pipelines. In contrast, \texttt{gemini-2.5-pro} appears strong on OWASP but collapses on \ov, illustrating the risk of relying on benchmarks alone. Efficient models like \texttt{o4-mini} occupy the low-cost, low-latency quadrant while maintaining consistent F1 across datasets, making them well-suited for pre-commit hooks and budget-conscious CI/CD use. These results highlight the need to align model selection with workflow demands and real-world reliability, not benchmark scores.  

\noindent \textbf{Core Challenges in Automated Adjudication.} Despite strong results, several challenges remain. {\it (i) Semantic reasoning vs. syntactic heuristics.} LLMs are better at understanding semantics than traditional static analyzers, but they still struggle to fully grasp developer intent, especially concerning custom sanitization or business logic. {\it (ii) Contextual blindness.} Providing the full context of a multi-file vulnerability is often infeasible. Our dataflow-based context extraction helps, but defining “sufficient context” remains an open research problem. {\it (iii) Evolving codebases.} As languages, frameworks, and vulnerability classes (CWEs) evolve, prompts and models can become stale. Continual evaluation and fine-tuning are essential for maintaining long-term robustness.

\section{Related Work}

\textbf{Automated Program Repair.} LLMs have recently been investigated for automated program repair, demonstrating promising capabilities in vulnerability remediation \cite{berabi2024deepcode, fu2023chatgpt, islam2024code, nong2024automated}. Prior studies show that models such as GPT-4 can generate plausible patches when combined with strategies like self-consistency, adaptive prompting, and ensemble filtering \cite{ahmed2023betterpatchingusingllm,fan2023automated,pearce2022examiningzeroshotvulnerabilityrepair}. More specialized frameworks, such as APPPatch \cite{nong2025appatchautomatedadaptiveprompting}, further advance this direction by isolating vulnerable code regions, synthesizing candidate fixes, and validating them automatically, achieving strong results on real-world CVEs. Collectively, these efforts underscore the growing potential of LLMs to bridge the gap between static vulnerability detection and fully automated remediation.

\vspace{1mm}

\noindent \textbf{Vulnerability Detection with LLMs.} LLMs have been studied as complements to static analysis, from reducing false positives in specific cases \cite{li2023assisting} and uncovering flaws missed by CodeQL \cite{ma2024lmsunderstandingcodesyntax,sun2023automaticcodesummarizationchatgpt} to broader evaluations showing strong but unstable performance \cite{lin2025large}, persistent challenges despite fine-tuning \cite{zhang2025benchmarking}, and steep drops on complex reasoning \cite{liu2024vuldetectbench}. Context has proven critical: interprocedural and commit-level information \cite{yildiz2025benchmarking}, context-rich benchmarks \cite{li2025everything}, natural language and contrastive prompts \cite{ceka2024can}, CWE-specific classifiers \cite{atiiq2024generalist}, and repository-level auditing with memory and validation \cite{guo2025repoaudit} all improve results. Advances in training further extend applicability, including reinforcement learning for reasoning \cite{simoni2025improving}, syntax- and semantic-aware pretraining \cite{li2025revisiting}, and synthesizing analyzers from patches to find long-latent Linux kernel bugs \cite{yang2025knighter}. Collectively, these works show LLMs advance vulnerability detection through enriched context, tailored prompting, and specialized training, while challenges in scalability and robustness remain.

\vspace{1mm}
\noindent \textbf{False Positive Mitigation.} Machine learning classifiers have long been used to distinguish true alerts from false positives \cite{10.1145/3510003.3510153,8730202,10.1145/3239235.3239523}, but despite success on controlled datasets, they depend heavily on labeled data and generalize poorly. To ease developer burden, other methods prioritize or cluster alerts based on severity, historical defects, or developer feedback \cite{10.1145/1349332.1349339,kremenek2003z,liu2018mining}, improving triage efficiency without discarding reports. More recently, LLMs have been integrated into static analysis to provide semantic reasoning beyond traditional heuristics. Prompting reduces false positives in use-before-initialization warnings \cite{li2023assisting}, while further work applies LLMs to control-flow inference, call-graph reasoning, and invariant detection \cite{ma2024lmsunderstandingcodesyntax,sun2023automaticcodesummarizationchatgpt,pmlr-v202-pei23a}. Building on this, LLM4SA \cite{wen2024automatically} automatically inspects static analysis warnings, achieving 81\% precision and 94\% recall, but remains limited to memory-related bugs with a uniform inspection strategy. LLM4FPM \cite{chen2024utilizing} improves context extraction through line-level slicing, though it evaluates only six CWEs, again focusing primarily on memory vulnerabilities. In contrast, \zf~ extends coverage to a wider variety of CWEs and evaluates performance across a greater number of state-of-the-art LLMs to provide deeper and more realistic insights. Unlike LLM4SA and LLM4FPM, which lack native SARIF support, our framework directly integrates SARIF, enabling seamless compatibility with any compliant tool and efficient adoption in CI/CD pipelines. Additionally, unlike prior efforts, our approach enriches the code context with CWE-specific details, ensuring that different alert types receive tailored treatment rather than a uniform procedure. Please note that a direct comparison between \zf~ and the aforementioned is not possible, as differences in datasets and experimental settings make reproduction infeasible.

\section{Conclusion}
False positives have long limited the usefulness of SAST, creating extra work and reducing developer trust. In this paper, we introduced \zf, a framework that combines static analysis with the reasoning ability of LLMs. By enriching alerts with dataflow context and CWE-specific knowledge, \zf\ allows LLMs to make more accurate, context-aware decisions about potential vulnerabilities.  Our results across benchmarks and real-world projects show that \zf\ reduces false positives while keeping precision high. We found that models built for reasoning perform better than those focused mainly on code generation or large context windows, and that CWE-specific prompting improves performance compared to general prompts. We also provide practical guidance for choosing models based on cost, latency, and accuracy needs.  By turning noisy alerts into reliable signals, \zf\ makes SAST a more trustworthy part of secure software development. It helps reduce wasted effort, restores confidence in automated security tools, and supports building more secure software at scale. In future work, we will extend \zf\ to more programming languages, improve context extraction for speed and scalability, and study its use in real CI/CD pipelines.  

\section{Data Availability}
To support reproducibility and enable further research, we provide a replication package containing the complete source code for all components of \zf. The package also includes scripts for setup and evaluation, the datasets used in our experiments, and the corresponding raw outputs. The replication package is publicly available at: \url{https://github.com/mhsniranmanesh/ZeroFalse}.


\newpage
\bibliographystyle{ACM-Reference-Format}
\bibliography{biblio}

\end{document}